\documentclass[ceqn, prd, showpacs, showkeys, prd, onecolumn, nofootinbib, amsmath, amsfonts, amssymb]{revtex4-2}

\usepackage{amssymb,amsmath,mathtools,xcolor,graphicx,xspace,colortbl,ragged2e,rotating} % 
\usepackage{textcomp}
\usepackage{amsfonts}  %% 010
\usepackage{amsmath}  %% 010
\usepackage{amssymb}  %% 100
\usepackage{boxedminipage}  %% 100
\usepackage{graphics}  %% 100
\usepackage{graphicx}  %% 100
\usepackage{ragged2e}  %% 100
\usepackage{tabulary}  %% 100
\usepackage{wrapfig}  %% 100
\usepackage{xcolor}  %% 200
\usepackage{varioref}  %% 1001
\graphicspath{{FRACTIONAL8_1_graphics/}{FRACTIONAL8_1_tcache/}{FRACTIONAL8_1_gcache/}}
\DeclareGraphicsExtensions{.pdf,.eps,.ps,.png,.jpg,.jpeg}

\begin{document}
\title{NEWTONIAN FRACTIONAL-DIMENSION GRAVITY AND THE MASS-DIMENSION FIELD EQUATION
}
\author{Gabriele U. Varieschi
}
\email[E-mail me at: ]{Gabriele.Varieschi@lmu.edu
}
\homepage[Visit: ]{https://gvarieschi.lmu.build
}
\affiliation{Department of Physics, Loyola Marymount University, Los Angeles, CA 90045, USA
}

\date{
\today
}

\begin{abstract}
We resume our analysis of Newtonian Fractional-Dimension Gravity (NFDG), an alternative gravitational model that does not require the dark matter (DM) paradigm. We add three more galaxies (NGC 6946, NGC 3198, NGC 2841) to the catalog of those studied with NFDG methods. Once again, NFDG can successfully reproduce the observed rotation curves by using a variable fractional dimension $D\left (R\right)$, as with the nine other galaxies previously studied with these methods. 
In addition, we introduce a mass-dimension field equation for our model, which is capable of deriving the fractional mass dimension $D_{m}\left(R \right)$ from a single equation, as opposed to the previous $D\left (R\right)$, which was obtained simply by matching the experimental rotational velocity data for each galaxy. While the NFDG predictions computed with this new  $D_{m}\left(R\right)$ dimension are not as accurate as those based on the original $D\left (R\right)$, they nevertheless confirm the validity of our fractional-dimension approach.
Three previously studied galaxies (NGC 7814, NGC 6503, NGC 3741) were analyzed again with these new methods, and their structure was confirmed to be free from any dark matter~components.

\end{abstract}
\maketitle

\section{\label{sect:intro}
Introduction
}

In the landscape of alternative gravitational theories, Newtonian Fractional-Dimension Gravity (NFDG) was first introduced in 2020--2021 with the first three papers (\cite{Varieschi:2020ioh,Varieschi:2020dnd,Varieschi:2020hvp}, hereafter papers I--III, respectively). Similar to other alternative models of gravity, the goal of NFDG was to explain galactic dynamics without using the dark matter (DM) paradigm. NFDG considers galaxies and possibly other astrophysical objects as fractal structures described by a fractional-dimension function $D\left (R\right)$, typically depending on the radial coordinate of the structure being studied and with real positive values (a Hausdorff-type fractal dimension, usually with $1 <D \lesssim 3$).

Papers I-III described how NFDG could be effectively applied to general structures with spherical or axial symmetries, as well as to the first three rotationally-supported galaxies (NGC 7814, NGC 6503, NGC 3741), without using any DM contribution. Paper IV \cite{Varieschi:2021rzk} introduced a relativistic version of the model, while paper V \cite{Varieschi:2022mid} analyzed \mbox{four additional} galaxies (NGC 5033, NGC 6674, NGC 5055, NGC 1090) in the context of the so-called External Field Effect (EFE). Finally, in the last NFDG paper (paper VI, \cite{Varieschi:2022nte}), we applied NFDG methods to the case of galaxies with little or no dark matter, such as the ultra-diffuse galaxies AGC 114905 and NGC 1052-DF2, and also briefly studied the Bullet Cluster merger (1E0657-56).
In all these cases, we were able to explain the dynamics of the aforementioned galaxies and of the Bullet Cluster merger with the NFDG model, without including any DM contributions.\protect\footnote{
	A general overview of Newtonian Fractional-Dimension Gravity can also be found in the NFDG website \cite{Varieschi:webpage}, together with updated analyses of all the galaxies studied with NFDG methods.
}

In our previous papers I-VI, possible connections with other well-established alternative gravities were introduced and analyzed in detail. In particular, we studied connections with modified Newtonian dynamics (MOND) \cite{Milgrom:1983ca,Milgrom:1983pn,Milgrom:1983zz,Famaey:2011kh}, conformal gravity (CG) \cite{Mannheim:1988dj,Mannheim:1992tr,Mannheim:2005bfa}, modified gravity (MOG) \cite{Moffat:2005si}, fractional gravity \cite{Calcagni:2011sz,Calcagni:2009kc,Calcagni:2010bj,Calcagni:2013yqa,Calcagni:2020ads,Calcagni:2021ipd,Calcagni:2021aap,Calcagni:2021mmj}, and other gravitational models. For general reviews on modified gravity and cosmology, dark matter, experimental tests of gravitational theories, etc., see also \cite{Clifton:2011jh,CANTATA:2021asi,Garrett:2010hd,Will:2014kxa}. 

The importance of the fractional/fractal approach in astrophysics can also be seen in other recent models, such as those by Giusti et al. \cite{Giusti:2020rul,Giusti:2020kcv}, Benetti et al.  \cite{Benetti:2023nrp,Benetti:2023vxy,Benetti:2023imt}, Llanes-Estrada \cite{Llanes-Estrada:2021hnt}, Moradpour et al. \cite{Moradpour:2024uqa}, and the $\kappa$-model by G. Pascoli \cite{Pascoli:2023dwn,Pascoli:2024med,Pascoli:2024dqg}, just to cite a few. All these studies are based on fractional calculus/fractional gravity and show connections to our NFDG formalism.

We also note that the possible physical origin of the fractional-dimension structure of the gravitational force is based on the description of astrophysical objects, such as galaxies, galaxy clusters, or others, as fractal objects \cite{bookBaryshev,bookMandelbrot,bookNottale,Calcagni:2016azd,bookFalconer,bookBarnsley}. Therefore, the fractional approach in astrophysics is currently relevant and might lead to further applications in gravitation and~cosmology.

In this paper, we will review and update in Section \ref{sect:revised} our NFDG standard methodology, which leads to the fractional dimension $D\left (R\right)$, by matching the experimental rotational velocity data for each galaxy under consideration. In Section \ref{sect:massdim}, we will introduce a simple mass-dimension field equation for our model, which will be used to derive the fractional mass dimension $D_{m}\left(R \right)$. In Section \ref{sect:galactic}, we will apply our computations to three new galaxies from the SPARC catalog \cite{Lelli:2016zqa,Li:2018tdo,Li:2022zms} (NGC 6946, NGC 3198, NGC 2841) and to three previously studied galaxies from the same catalog (NGC 7814, NGC 6503, NGC 3741). In particular, for each galaxy, we will compare the results obtained by using the two different fractional dimensions $D\left (R\right)$ and $D_{m}\left(R \right)$. Final conclusions will be reported in Section  \ref{sect:conclusion}.

\section{\label{sect:revised} Revised NFDG computations
}

NFDG was derived from a heuristic extension of Gauss's law for gravitation to a lower-dimensional space--time $D +1$, where $D \leq 3$  can be considered a non-integer space dimension \cite{Varieschi:2020ioh,Varieschi:2020dnd,Varieschi:2020hvp}. The NFDG gravitational potential for a point mass $m$ placed at the origin is as follows\cite{Varieschi:2022mid}:
\begin{gather}\Phi _{NFDG}(r) = -\frac{2\pi ^{1 -D/2}\Gamma (D/2)\ Gm}{\left (D -2\right)l_{0}r^{D -2}}\ ;\ D \neq 2 \label{eq2.1} \\
	\Phi _{NFDG}\left (r\right) =\frac{2\ Gm}{l_{0}}\ln r\ ;\ D =2 . \nonumber \end{gather}

Here, $G$ is the gravitational constant, the radial coordinate $r$ is considered to be dimensionless, and $l_{0}$ is an appropriate scale length, usually introduced in fractional gravity models for dimensional correctness.

The generalization of Gauss's law to non-integer dimension spaces was established by using dimensional regularization techniques commonly used in quantum field theories and was also connected to Weyl's fractional integrals, thus establishing a direct relationship between fractal space--time and fractional calculus (see Ref. \cite{Varieschi:2020ioh} for full details). In addition, the NFDG gravitational potential satisfies a generalized fractional Poisson equation (see Equation (22) in Ref. \cite{Varieschi:2020ioh}), based on the generalized form of the D-dimensional Laplace operator. Related solutions to the D-dimensional Laplace equation in spherical coordinates and corresponding multipole expansions were also presented in our first paper, as well as general theorems for spherically symmetric mass distributions in D-dimensional spaces (see Appendix A in Ref. \cite{Varieschi:2020ioh}). 

NFDG formulas for axially symmetric mass distribution were introduced in paper II \cite{Varieschi:2020dnd} and then refined in subsequent papers III \cite{Varieschi:2020hvp} and V \cite{Varieschi:2022mid}. In paper IV \cite{Varieschi:2021rzk}, we reviewed and expanded the Euler--Lagrange equations for scalar fields in fractional D-dimension spaces (rectangular, spherical, and cylindrical coordinates) and also introduced a relativistic version of NFDG based on a modified Hilbert action. Therefore, we feel that NFDG is strongly based on the mathematical theory of spaces with non-integer dimension  \cite{doi:10.1063/1.523395,1987JPhA...20.3861S,Palmer2004}, and it is not an ad hoc model just capable of fitting the experimental data.

In our previous NFDG papers I--VI, we also linked directly our scale length $l_{0}$ to the MOND acceleration constant $a_{0} \simeq 1.2 \times 10^{ -10}\:\mbox{m}\thinspace \mbox{s}^{ -2}$, namely $a_{0} \approx GM/l_{0}^{2}$ for a galaxy of mass $M$. In this paper, we will not use this direct connection anymore since our NFDG computations of the gravitational potential and related circular velocities are actually independent of the choice of the $l_{0}$ parameter (see Appendix A of Ref. \cite{Varieschi:2022mid} for full details). However, we have shown in paper I \cite{Varieschi:2020ioh} that, by assuming the above connection between $l_{0}$ and $a_{0}$, NFDG can easily recover the fundamental results of the deep-MOND regime \cite{Milgrom:1983ca,Milgrom:1983pn,Milgrom:1983zz}, including the flat rotational speed $V_{f}=\sqrt[4]{GMa_{0}}$, by setting $D=2$ in the expressions for the NFDG gravitational field.

The NFDG potential in Equation (\ref{eq2.1}) was then generalized to extended source mass distributions \cite{Varieschi:2022mid}, such as spherically symmetric and axially symmetric distributions, which are used to model the three main components of the galactic baryonic matter: the spherical bulge mass distribution (if present) and the cylindrical stellar disk and gas distributions. 

In this work, we continue using the light-to-mass ratios as in our previous papers: $\Upsilon _{disk} \simeq 0.50\ M_{ \odot }/L_{ \odot }$, $\Upsilon _{gas} \simeq 1.33\ M_{ \odot }/L_{ \odot }$ (this value for $\Upsilon _{gas}$ includes also the helium gas contribution), and $\Upsilon _{bulge} \simeq 0.70\ M_{ \odot }/L_{ \odot }$. These light-to-mass ratios were used to convert SPARC surface luminosity data into corresponding surface mass distributions: $\Sigma_{disk}\left (R\right)$, $\Sigma_{gas}\left (R\right)$, and $\Sigma_{bulge}\left (R\right)$. The bulge surface mass distribution can then be turned into a more appropriate spherically symmetric mass distribution $\rho_{bulge} \left (r\right)$ by applying Equation (1.79) in Ref. \cite{bookBinney}.
We also assume that the space dimension $D$ is a function of the field point coordinates, while neglecting its space derivatives in the computation of the NFDG gravitational field:
\begin{equation}\mathbf{g}_{NFDG}\left (R\right) = -\frac{1}{l_{0}}\frac{d\Phi _{NFDG}\left (R\right)}{dR}\widehat{\mathbf{R}}. \label{eq2.2}
\end{equation}

The gravitational field is computed in the galactic disk plane and expressed as a function of the dimensionless radial coordinate $R$ in the same plane (the scale length $l_{0}$ is included into the definition of $R$, for dimensional correctness). Similarly, the variable dimension $D =D\left (R\right)$ is considered a function of the same radial coordinate and will characterize each particular galaxy studied with NFDG methods.

The circular velocity, for all baryonic matter rotating in the main galactic plane, is then obtained from Equation (\ref{eq2.2}) as
\begin{equation}v_{circ}\left (R\right) =\sqrt{l_{0}R\left \vert \mathbf{g}_{NFDG}\left (R\right)\right \vert }, \label{eq2.3}
\end{equation}
with the NFDG field computed by using the variable dimension function $D =D\left (R\right)$ mentioned above. 
Full details about the NFDG computations can be found in our papers I-V, with the latest version described in Appendix A of paper V \cite{Varieschi:2022mid}.

A key element of this computation is the series expansion of the $1/r^{D-2}$ term in the first line of Equation (\ref{eq2.1}).
Rewriting this term by using the distance between the field point $\mathbf{x}$ and the source point $\mathbf{x}^{ \prime }$, the Euler kernel $1/\left \vert \mathbf{x} -\mathbf{x}^{ \prime }\right \vert ^{D -2}$ can be expanded for $D >1$, $D \neq 2$, using rescaled spherical coordinates \cite{2012JPhA...45n5206C}:
\begin{equation}\frac{1}{\left \vert \mathbf{x} -\mathbf{x}^{ \prime }\right \vert ^{D -2}} =\sum \limits _{l =0}^{\infty }\frac{r_{ <}^{l}}{r_{ >}^{l +D -2}}C_{l}^{\left (\frac{D}{2} -1\right)}\left (\cos \gamma \right), \label{eq2.4}
\end{equation}
where $r_{ <}$ ($r_{ >}$) is the smaller (larger) of $r$ and $r^{ \prime }$, $\gamma $ is the angle between the unit vectors $\widehat{\mathbf{r}}$ and $\widehat{\mathbf{r}}^{ \prime }$, and $C_{l}^{\left (\lambda \right)}\left (x\right)$ denotes Gegenbauer polynomials \cite{NIST:DLMF}.

Equation (\ref{eq2.4}) can then be used to model both spherically symmetric galactic components (bulge) and cylindrical components (disk and gas), by using appropriate choices of coordinates and angles as described in detail in paper V. After combining all these elements into a single formula for the cylindrical/spherical NFDG potentials and fields, very cumbersome final expressions are obtained (see Equations (A9)--(A13) in Appendix A of Ref. \cite{Varieschi:2022mid}).

However, the series expansions due to the presence of the Euler kernel in all final formulas are seen to converge rapidly, thus allowing for the inclusion of only the first few terms. In our previous papers I-VI, we summed the first six non-zero terms of the expansions  ($l =0,2,4,6,8,10$, in Equation (\ref{eq2.4})). In this paper, we simplified the summations by including only the first four non-zero terms ($l =0,2,4,6$, in Equation (\ref{eq2.4})). We checked that the difference in the final results, computed with the old and the new method, typically amounts to less than one percent. This simplification allowed a considerable reduction in the time needed to run our Mathematica programs on standard computing machines.

Even with this simplification in the series expansions, our final formulas for the NFDG potentials/fields/circular velocities are still complicated expressions, which are not suitable for a direct fit to the SPARC experimental data. Therefore, we continued using our previous strategy of finding the fractional dimension function $D\left (R\right)$ by matching the experimental data with the NFDG fits for a fixed number of points, over the range of radial distances. In past papers I--VI, we considered 100 equally spaced points over the radial range and found at each point the best value of $D\left (R\right)$, which allowed our NFDG formulas to match the experimental circular velocity. To further reduce computing times, in this paper, we elected to limit our fits to 50 points over the radial range. These simplifications in our main procedure do not seem to affect the quality of our final results; so, we will use them as our standards also in future computations.

However, it should be noted that despite all the simplifications introduced in our computations and described above, the practical time needed to fully analyze one single galaxy with our Mathematica routines is approximately still 3--4 days on a standard computer. This element and other factors limit our ability to produce NFDG analyses for many more (or for all the 175) galaxies in the SPARC catalog. We will continue to analyze additional galaxies in upcoming publications or post updated results directly in our NFDG website~\cite{Varieschi:webpage}.

\section{\label{sect:massdim}NFDG mass-dimension field equation
}

As mentioned in the preceding section, the fractional-dimension $D =D\left (R\right)$ is not derived from a single equation but rather obtained by matching our NFDG formulas to the experimental data for each galaxy. This has raised concerns, in relation to our previous papers, about the ability of NFDG to derive the fractional dimension function from some fundamental field equation. It is obvious that this ``field equation'' would need to determine the fractional dimension of an astrophysical object, i.e., its fractal nature, solely from the known distribution of baryonic matter in the same object and without any DM~contributions.

Therefore, we need to determine a new mass-dimension function  $D_{m}\left(R\right)$ directly from the bulge/disk/gas mass distributions of each galaxy being considered. In Appendix C of our paper V \cite{Varieschi:2022mid}, we considered a simplified method to compute the mass dimension and applied it to the case of NGC 5055. In this section, we expand on this previous method and determine a more general NFDG mass-dimension field equation.

The mass dimension of a fractal system describes how the mass of a structure scales with its size. It essentially quantifies how much space a fractal object fills, reflecting its density and complexity. A higher mass fractal dimension implies a more space-filling structure. Considering an isotropic fractal material, the definition of the mass dimension $D_{m}$ is the following \cite{bookTarasov,Tarasov:2014fda,TARASOV2015360}: $M_{D}\left(W_{B} \right)=M_{0}{\left(\frac{R}{R_{0}} \right)}^{D_{m}}$. Here, $M_{D}$ represents the mass of a ball $W_{B}$ of radius $R$ of the fractal material, while $R_{0}$ is a characteristic scale length of the fractal material. $M_{0}$ is the mass of the ball of radius $R_{0}$. For $D_{m}=3$, we obtain the customary result that the mass of a ball of uniform density scales like the cube of the radius~$R$.

However, galactic fractal structures are typically more shaped like disks than spherical objects, and their mass distributions are described using a total surface mass distribution, $\Sigma_{tot} \left (R\right) =\Sigma _{bulge}\left (R\right) +\Sigma _{disk}\left (R\right)+\Sigma _{gas}\left (R\right)$, for the three main components (see Appendix A of Ref. \cite{Varieschi:2022mid} for details).

The bulge mass distribution is obviously spherically symmetric and should be described by a spherical mass distribution $\rho_{bulge} \left (r\right)$. However, as explained in Section \ref{sect:revised} above, the original SPARC luminosity data for the bulge are given in terms of its surface luminosity, which is then converted directly into the bulge surface mass density $\Sigma_{bulge}\left (R\right)$ by using the light-to-mass factor $\Upsilon _{bulge} \simeq 0.70\ M_{ \odot }/L_{ \odot }$ mentioned in Section \ref{sect:revised}. Therefore, we can combine the bulge surface mass distribution with the disk and gas surface mass distributions and obtain the total $\Sigma_{tot} \left (R\right)$, by treating all the three mass distributions as axially symmetric.

Moreover, the bulge distribution is sometimes absent from certain galaxies in the SPARC database (for example, only three out of the six galaxies analyzed in Section \ref{sect:galactic} possess a central bulge) and, when the bulge is present, it usually dominates at low radial distances, while the axially symmetric disk and gas components are usually more prominent at larger distances where the flattening of the rotation curves is observed.

All the above considerations suggest adjusting the definition of the mass dimension $D_{m}$ as follows:
\begin{equation}M\left(R \right) = M_{0}{\left(\frac{R}{R_{0}} \right)}^{D_{m}(R)-1}, \label{eq3.1}
\end{equation}
so that the mass of a surface distribution will scale as $R^2$ for $D_{m}=3$. We also consider the mass dimension $D_{m}$ as a function of the radial coordinate $R$.

We can then differentiate the last expression:
\begin{equation}dM\left(R \right) = \frac{M_{0}}{R_{0}}\left(\frac{R}{R_{0}} \right)^{\left(D_{m}\left(R \right)-2 \right)}\left[\left(D_{m}\left(R \right)-1 \right)dR+R\ln\left(\frac{R}{R_{0}} \right)dD_{m}\left(R \right)  \right] \label{eq3.2}
\end{equation}
and compare it to the equivalent NFDG expression due to the total surface mass distribution $\tilde{\Sigma}_{tot}$ in a D-dimensional space:\protect\footnote{This follows from the fundamental Equation (1) in paper IV \cite{Varieschi:2021rzk} for spherically symmetric functions, originally introduced by Stillinger \cite{doi:10.1063/1.523395} and Svozil \cite{1987JPhA...20.3861S}, adapted to an axially symmetric function $f=f(R)$ in a $D-$dimensional metric space $\chi$, i.e., \begin{equation}\int _{\chi }fd\mu _{H} =\frac{2\pi ^{(D-1)/2}}{\Gamma ((D-1)/2)}\int _{0}^{\infty }f(R)R^{D -2}dR, \nonumber
	\end{equation}
	where $\mu _{H}$ denotes an appropriate Hausdorff measure over the space. See Ref. \cite{Varieschi:2021rzk} for more details.}
\begin{equation}dM\left(R \right) = \frac{2\pi^{\left(\frac{D_{m}\left(R \right)-1}{2} \right)}}{\Gamma\left(\frac{D_{m}\left(R \right)-1}{2}  \right)} \tilde{\Sigma}_{tot}\left(\frac{R}{R_{0}} \right)\left(\frac{R}{R_{0}} \right)^{\left(D_{m}\left(R \right)-2 \right)}\frac{dR}{R_{0}} . \label{eq3.3}
\end{equation}

In the previous equations, the quantity $R_{0}$ acts as a scale length for our fractional-dimension structures, equivalent to the original NFDG scale length $l_{0}$ in Equations (\ref{eq2.1})--(\ref{eq2.3}). In addition, the total surface mass density $\tilde{\Sigma}_{tot}(R/R_{0})$ in the last equation is a ``rescaled'' mass density, i.e., $\tilde{\Sigma}_{tot}(R/R_{0})=\Sigma_{tot}(R) \: R_{0}^{2}$ with dimensions of mass; thus, Equation (\ref{eq3.3}) is dimensionally correct. Combining together the last two equations and simplifying, we~obtain
\begin{equation} D_{m}\left(R \right)-1 +R\ln\left(\frac{R}{R_{0}} \right)\frac{dD_{m}\left(R \right)}{dR}   = \frac{2\pi^{\left(\frac{D_{m}\left(R \right)-1}{2} \right)}}{\Gamma\left(\frac{D_{m}\left(R \right)-1}{2}  \right)} \frac{\tilde{\Sigma}_{tot}\left({R}/{R_{0}} \right)}{M_{0}}, \label{eq3.4}
\end{equation}
which will be considered the NFDG mass-dimension field equation and will be solved numerically in Section \ref{sect:galactic} for several different galaxies.

Equation (\ref{eq3.4}) is a first-order non-linear differential equation which allows deriving the mass-dimension function $D_{m}\left(R \right)$ from a more foundational basis, as opposed to the ``heuristic'' dimension function $D\left(R \right)$ used in Section \ref{sect:revised}. The mass-dimension function $D_{m}\left(R \right)$ can be considered the main ``field'' in NFDG, since it determines the dynamics of the fractal astrophysical structure, when used together with the main NFDG \mbox{Equations (\ref{eq2.1})--(\ref{eq2.3}).}

Equation (\ref{eq3.4}) is an ordinary differential equation, since we are not including any time dependence of the dimension function, and we only use the axial radial coordinate $R$ in our analysis. In the original theory of isotropic fractal materials \cite{bookTarasov,Tarasov:2014fda,TARASOV2015360}, the scale radius $R_{0}$ represents the characteristic scale of the fractal medium at which fractal behavior starts being observed, and $M_{0}$ is the corresponding mass of a ball of radius $R_{0}$. In our NFDG analysis, we will simply consider $R_{0}$ and $M_{0}$ as free parameters of our model. The only physical input in Equation (\ref{eq3.4}) is represented by the (rescaled) total mass density function $\tilde{\Sigma}_{tot}(R/R_{0})$, which will depend on the total baryonic mass distribution of the galaxy being~studied.

In the next section, we will study three new galaxies in the SPARC catalog, combining together the NFDG methods outlined in Section \ref{sect:revised} and  Section \ref{sect:massdim}. This will also be the first test of our main field Equation (\ref{eq3.4}) above.

\section{\label{sect:galactic}NFDG and galactic data fitting
}

In the following subsections, we will apply NFDG to three notable examples of rotationally-supported galaxies from the SPARC database: NGC 6946, NGC 3198, and NGC 2841.  Luminosity data for the bulge, disk, and gas components of these three galaxies are available in the SPARC database \cite{Lelli:2016zqa,Li:2018tdo,Li:2022zms}, while supplemental information was obtained from the database administrator \cite{Lelli:2024pri}. The choice of these galaxies is due to the fact that they were used as the main examples in several recent papers on the subject \cite{Famaey:2011kh,Li:2018tdo,Chae:2020omu,Li:2020iib,Li:2022zms}. We will also reconsider three other galaxies (NGC 7814, NGC 6503, NGC 3741), which were analyzed in previous papers.

\subsection{\label{sectiongalacticone}NGC 6946}
\label{sectiongalacticone}

We start with the case of NGC 6946, an intermediate spiral galaxy in the Virgo Supercluster, approximately 25 million light-years away. This galaxy is sometimes called the ``Fireworks Galaxy'' due to its face-on aspect and prominent and continuous spiral arms. It has a small bright central nucleus, surrounded by a strong stellar disk  and gas components extending outward in space. This is consistent with the SPARC luminosity data, which indicate a strong bulge component up to about 1--2 $\mbox{kpc}$, beyond which, the stellar disk dominates up to the largest radial distances, with a relatively less strong gas component. This galaxy was also prominently featured in Ref. \cite{Famaey:2011kh}, in the general context of MOND observational phenomenology and with particular emphasis on the so-called Renzo's rule \cite{Sancisi:2003xt}, i.e., the strong correlation between features in the luminosity profile and corresponding features in the galactic rotation curve, which cannot be easily explained by the DM paradigm.

SPARC data for this galaxy also include the following: distance $D =\left (5.52 \pm 1.66\right)\ \mbox{Mpc}$, disk scale length $R_{d} =2.44\ \mbox{kpc} =7.53 \times 10^{19}\ \mbox{m}$, asymptotically flat rotation velocity $V_{f} =\left(158.9 \pm 10.9\right)\ \mbox{km/s}$. Using the SPARC luminosity data for the three main components, we computed the corresponding volume and surface mass distributions, following the procedure discussed in Appendix A of paper V. Integrating the mass distributions, we obtained the following galactic masses: $M_{bulge} =3.53 \times 10^{39}\ \mbox{kg}$, $M_{disk} =6.16 \times 10^{40}\ \mbox{kg}$, $M_{gas} =1.52 \times 10^{40}\ \mbox{kg}$, and $M_{total} =8.03 \times 10^{40}\ \mbox{kg}$.

\begin{figure}\centering 
	\setlength\fboxrule{0in}\setlength\fboxsep{0.1in}\fcolorbox[HTML]{000000}{FFFFFF}{\includegraphics[width=6.99in, height=8.728805970149253in,]{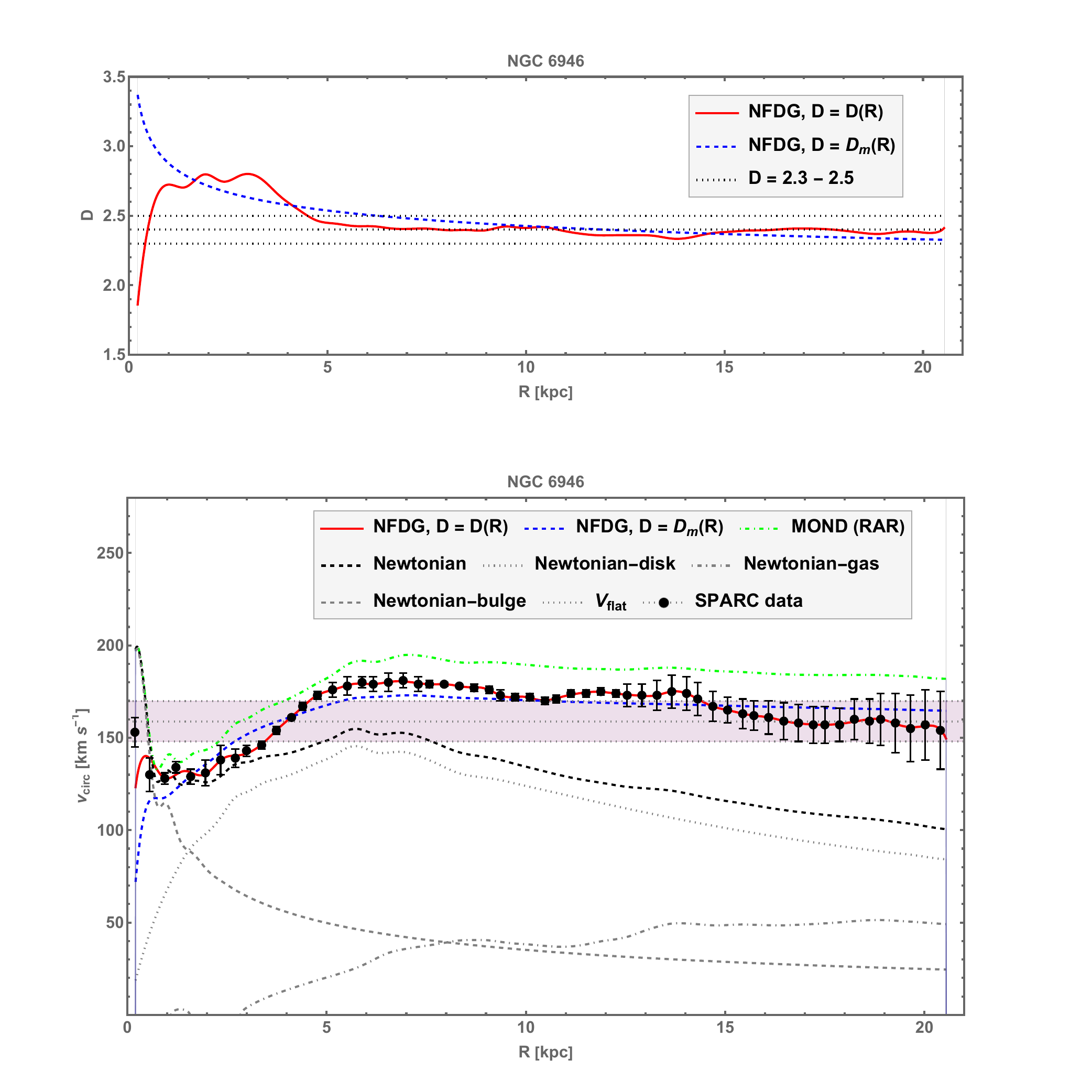}}
	\caption{NFDG results for NGC 6946.
		Top panel: NFDG variable dimension $D\left (R\right )$, based directly on SPARC data (red-solid curve), compared with NFDG mass-dimension $D_{m}\left(R\right)$ (blue-dashed curve), and fixed values $D =2.3-2.5$ (black-dotted lines). Bottom panel: NFDG rotation curves (circular velocity vs. radial distance) compared to the original SPARC data (black circles with error bars). The NFDG best fit for the variable dimension $D\left (R\right )$ is shown by the red-solid line, while the NFDG fit for the mass-dimension $D_{m}\left(R\right)$ is shown by the blue-dashed curve. Also shown: MOND prediction based on the general RAR (green, dot-dashed), Newtonian rotation curves (different components - gray lines, total - black dashed line), and asymptotic flat velocity band (horizontal gray band).}
	\label{figure:NGC6946}
\end{figure}

The NFDG results for this galaxy are shown in Figure \ref{figure:NGC6946}. As performed in previous papers, we measure the radial distance $R$ in \textrm{kiloparsec} and rotation circular velocities $v_{circ}$ in $\mbox{km}\ \mbox{s}^{ -1}$. The computations follow the methods detailed in Sections \ref{sect:revised}--\ref{sect:massdim}, with radial limits for the computation set at $R_{\min } =0.21\ \mbox{kpc}
$ and $R_{\max } =20.5\ \mbox{kpc}$. These limits are shown in all figures as vertical thin-gray lines.

As explained in Section \ref{sect:revised}, the fundamental NFDG result in this figure is the variable dimension $D\left (R\right)$ shown in the top panel by the red-solid curve. As in our previous papers I-VI, this was obtained by interpolating the experimental SPARC data for the circular velocities (black circles with error bars in the bottom panel) and by deriving from them the equivalent observed gravitational field $g_{obs}\left (R\right)$, based exclusively on SPARC experimental data. We then assumed that this observed field is the same as $g_{NFDG}\left (R\right)$ in Equation (\ref{eq2.2}), obtained from the NFDG gravitational potential, as described in Section \ref{sect:revised}.

At each field point, the NFDG potential and related gravitational field are also considered as functions of the variable dimension $D\left (R\right)$, i.e., $g_{NFDG}\left (R,D\left (R\right)\right)$. The computational range $\left (R_{\min },R_{\max }\right)$ was divided into equal sub-intervals with related radial distances $R_{i} =R_{\min } +i\genfrac{(}{)}{}{}{R_{\max } -R_{\min }}{50},\ \ i =0, \dots,50$, and for each of these $R_{i}$ points the following equation was solved numerically:
\begin{equation}g_{NFDG}\left (R_{i},D\left (R_{i}\right)\right) =g_{obs}\left (R_{i}\right) . \label{eq4.1}
\end{equation}

This leads to the determination of the corresponding values of the variable dimension $D_{i} \equiv D\left (R_{i}\right),\ i =1, \dots,50$.

By interpolation of this set of $\left (R_{i},D_{i}\right)$ points, the fundamental NFDG red-solid variable dimension $D\left (R\right)$ curve in the top panel of Figure \ref{figure:NGC6946} is obtained. As a final check of our procedure, at each radial point $R_{i}$ we recompute the NFDG circular velocities using the $D\left (R\right)$ function and Equation (\ref{eq2.3}). This yields the (almost) perfect NFDG fit to the SPARC experimental data shown by the red-solid curve in the bottom panel of Figure \ref{figure:NGC6946}. This perfect agreement is expected, since, at each point, we select the appropriate value of the variable dimension $D\left (R_{i}\right)$, which allows matching the experimental value $g_{obs}\left (R_{i}\right)$ to the predicted NFDG value $g_{NFDG}\left (R_{i},D\left (R_{i}\right)\right)$, following Equation (\ref{eq4.1}). 

As detailed in Section \ref{sect:massdim}, an alternative method can be used to derive the NFDG mass-dimension function $D_{m}\left(R\right)$ directly from the bulge/disk/gas mass distributions of NGC 6946, by using the field Equation (\ref{eq3.4}). This yields the blue-dashed curve in the top panel of Figure \ref{figure:NGC6946} for $D_{m}\left(R\right)$, and the corresponding blue-dashed curve for the circular velocity in the bottom panel of the same figure.

To obtain $D_{m}\left(R\right)$, we used the ``NonlinearModelFit'' function in our Mathematica \cite{Mathematica2024} routines with the model defined as the differential equation in Equation (\ref{eq3.4}). The reference radial length $R_{0}$ and related reference mass $M_{0}$ were left as free parameters, while a required initial condition for the differential equation was set by choosing one of the last SPARC data points for this galaxy, at distance $R_{data}$, and with circular velocity $V_{data}$ close to the asymptotically flat rotation velocity $V_{f}$ mentioned above. The initial condition was then set as 
\begin{equation}D_{m}\left(R_{data} \right)=D\left(R_{data} \right), \label{eq4.2}
\end{equation}
i.e., we set the value of the mass dimension $D_m$ at this reference distance $R_{data}$ to be equal to the value of the known fractional dimension $D$ at the same distance.

Again, the results of this procedure are shown in both panels of Figure \ref{figure:NGC6946} by the blue-dashed curves. Although the blue-dashed fit to the circular velocity data in the bottom panel is not as accurate as the one obtained with the previous method (red-solid curve), it is still able to capture the overall trend of the data and even to describe some of the features in the observed data. Similarly, the blue-dashed $D_{m}\left(R\right)$ curve in the top panel appears to describe well the overall evolution of the variable dimension, although it does not show the details of the red-solid $D\left(R\right)$ function.

We also tried modifying the value of $R_{data}$ in the initial condition mentioned in \mbox{Equation (\ref{eq4.2})} and changing other parameters of the computation, but the results for $D_{m}\left(R\right)$ and related fits were not much affected. We can conclude that our field Equation (\ref{eq3.4}) for $D_{m}\left(R\right)$ is effective in predicting the rotational velocity pattern of NGC 6946 without any DM contributions, although it is not as accurate as the method based on the original $D\left (R\right)$. However, both methods confirm the validity of our fractional-dimension approach to galactic dynamics.

In the bottom panel of Figure \ref{figure:NGC6946}, we also show a MOND prediction (green, dot-dashed curve) based on the general Radial Acceleration Relation (RAR) formula \cite{McGaugh:2016leg,Lelli:2017vgz}:
\begin{equation}g_{obs} =\frac{g_{bar}}{1 -e^{ -\sqrt{g_{bar}/g_{\dag }}}}, \label{eq4.3}
\end{equation}
where $g_{\dag } =1.20 \times 10^{ -10}\ \mbox{}\ \mbox{m}\thinspace \mbox{s}^{ -2}$ is equivalent to the MOND acceleration scale $a_{0}$.

We note that this MOND (RAR) fit is performed here by using directly the formula in Equation (\ref{eq4.3}) and the SPARC data for $g_{bar}$ without any further adjustment of the parameters, as opposed to the individual analyses of SPARC galaxies with the RAR performed in \mbox{Ref. \cite{Li:2018tdo},} where additional quantities (mass-to-light ratios, galaxy distance, and disk inclination of each galaxy) were used as free parameters to improve the fitting procedure.

As seen from the green dot-dashed curve in the bottom panel of Figure \ref{figure:NGC6946}, the MOND (RAR) fit without adjustment of the parameters does not yield a perfect fit, but it is able to approximately describe the overall pattern of the empirical data. In the bottom panel of Figure \ref{figure:NGC6946}, we also detail the Newtonian rotation curves (different components---gray lines; total---black dashed line) and the asymptotic flat velocity band (horizontal gray band), based on the $V_f$ data for this galaxy reported above. These Newtonian curves are the result of the direct interpolation of the SPARC data.

As a general comment about the NFDG results, we point out again that our main outcome is shown by the red-solid variable dimension curve in the top panel of Figure \ref{figure:NGC6946}: if this galaxy were to behave as a fractal structure with Hausdorff fractional dimension following the $D\left (R\right)$ function, then the related NFDG red-solid rotation curve would match the experimental data, without needing any form of dark matter. The $D\left (R\right)$ curve for NGC 6946 seems to be increasing toward standard $D \approx 3$ values at lower radial distances, where the spherical bulge is more dominant, while decreasing at larger distances and approaching an almost constant $D \approx 2.4$ value at the largest radii, where the asymptotic flat velocity regime takes place. Similar behavior was seen for other galaxies studied in the past and also for those that will be described in the following subsections.

A similar trend is also shown by the mass-dimension function $D_{m}\left (R\right)$ introduced for the first time in this paper. The blue-dashed curve in the top panel of Figure \ref{figure:NGC6946} remains in the range  $D \approx 2.3-2.5$ at larger radial distances, while increasing at lower distances toward standard $D \approx 3$. As noted in previous papers, our NFDG computations and fits become less reliable at the lowest radial distances, due to convergence problems with our series expansion formulas and the lack of reliable mass density data near the galactic center. 

\subsection{NGC 3198}
\label{sectiongalactictwo}

As our second galaxy, we consider here NGC 3198, a barred spiral galaxy in the constellation Ursa Major, which is also part of the Virgo Supercluster and approximately \mbox{47 million} light-years away. This galaxy does not possess a central bulge, only a dominant stellar disk component, together with a relatively less strong gas component. NGC 3198 and its dynamics are discussed in several papers of the SPARC group \cite{2002ARA&A..40..263S,Li:2018tdo,Chae:2020omu,Li:2022zms}, and this galaxy was sometimes also regarded as a problematic case for MOND \cite{Gentile:2010xt,Gentile:2013tfa}.

\begin{figure}\centering 
	\setlength\fboxrule{0in}\setlength\fboxsep{0.1in}\fcolorbox[HTML]{000000}{FFFFFF}{\includegraphics[ width=6.99in, height=8.728805970149253in,]{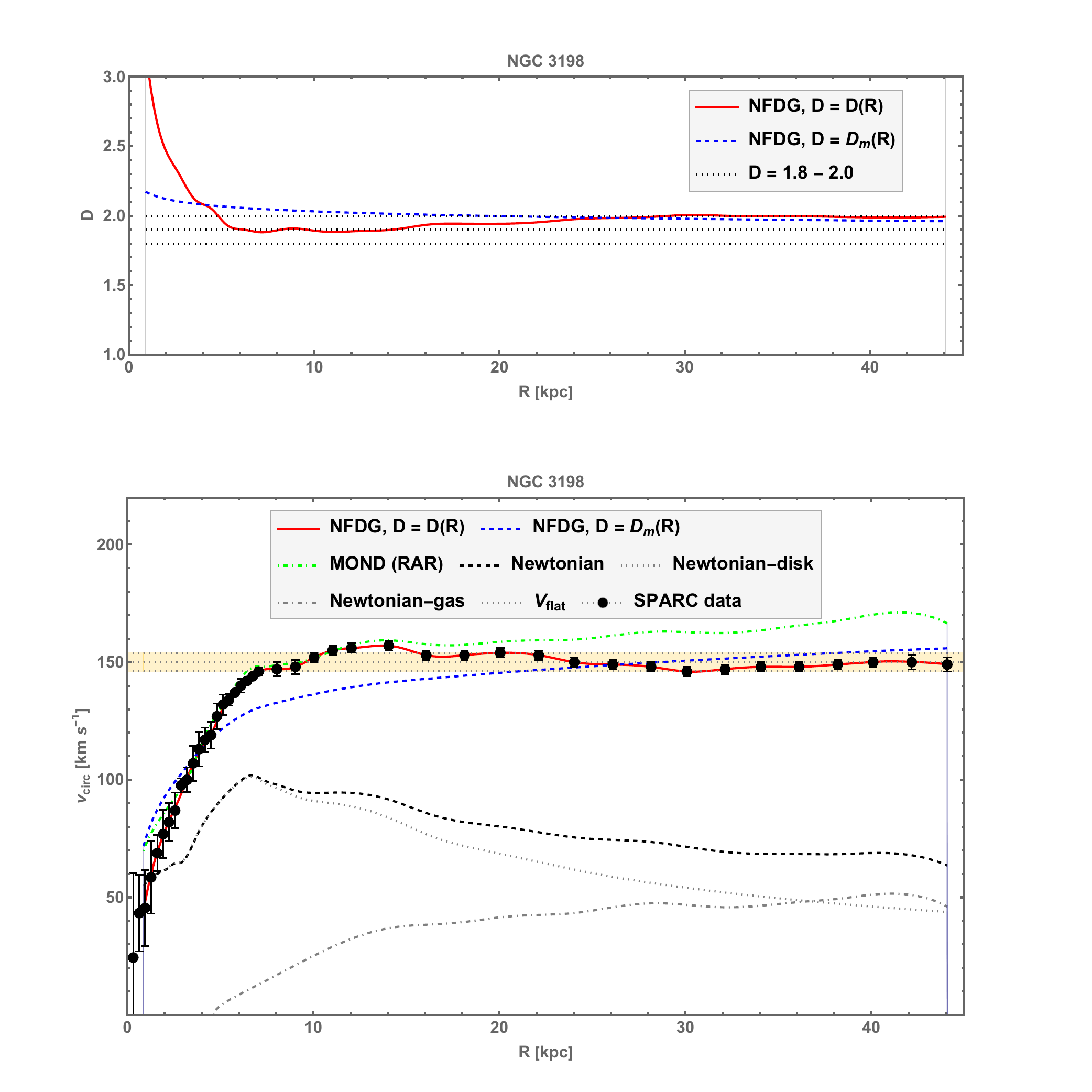}
	}
	\caption{NFDG results for NGC 3198.
		Top panel: NFDG variable dimension $D\left (R\right )$, based directly on SPARC data (red-solid curve), compared with NFDG mass-dimension $D_{m}\left(R\right)$ (blue-dashed curve), and fixed values $D =1.8-2.0$ (black-dotted lines). Bottom panel: NFDG rotation curves (circular velocity vs. radial distance) compared to the original SPARC data (black circles with error bars). The NFDG best fit for the variable dimension $D\left (R\right )$ is shown by the red-solid line, while the NFDG fit for the mass-dimension $D_{m}\left(R\right)$ is shown by the blue-dashed curve. Also shown: MOND prediction based on the general RAR (green, dot-dashed), Newtonian rotation curves (different components - gray lines, total - black dashed line), and asymptotic flat velocity band (horizontal gray band).}
	\label{figure:NGC3198}
\end{figure}

SPARC data for NGC 3198 include the following: distance $D =\left (13.80 \pm 1.40\right)\ \mbox{Mpc}$, disk scale length $R_{d} =3.14\ \mbox{kpc} =9.69 \times 10^{19}\ \mbox{m}$, asymptotically flat rotation velocity $V_{f} =\left (150.1 \pm 3.9\right)\ \mbox{km/s}$. By integrating the SPARC mass distributions, we computed the following galactic masses: $M_{disk} =3.77 \times 10^{40}\ \mbox{kg}$, $M_{gas} =2.83 \times 10^{40}\ \mbox{kg}$, and \mbox{$M_{total} =6.60 \times 10^{40}\ \mbox{kg}$.} The NFDG results for this galaxy are illustrated in Figure \ref{figure:NGC3198}, with the radial limits set at $R_{\min } =0.87 \ \mbox{kpc}
$ and $R_{\max } =44.1 \ \mbox{kpc}$ (vertical thin-gray lines in the~figure).

The variable dimension $D\left (R\right)$ and the mass dimension $D_{m}\left (R\right)$ are shown in the top panel by the red-solid and blue-dashed curves, respectively, and were obtained with the same procedures outlined in Sections \ref{sect:revised}--\ref{sect:massdim} and in subsection \ref{sectiongalacticone} above.  The corresponding NFDG circular velocity fits are shown in the bottom panel of the same figure: the red-solid curve produces a practically perfect fit to the SPARC data, while the blue-dashed curve based on our new field Equation (\ref{eq3.4}) produces an approximate fit, which still captures the overall pattern of the data. Similarly, the MOND (RAR) fit, based on Equation (\ref{eq4.3}), without any parameter adjustments, can only approximate the experimental data.

The general analysis for this galaxy is similar to the previous one for NGC 6946, with some differences: the absence of a spherical bulge component is consistent with lower values of both $D\left (R\right)-D_{m}\left (R\right)\approx 2.0$ at large radial distances. This behavior is typical of flat-disk mass distributions in NFDG, where the fractal dimension asymptotically approaches $D \approx 2$ at the largest radial distances. This was seen also in NGC 6503 \cite{Varieschi:2022mid}, which will be revisited in Section \ref{sectiongalacticfour}.
At the lowest radial distances, the variable dimension becomes close to the standard value $D \approx 3$, which is expected, as purely Newtonian predictions are effective at small distances from the galactic center.

\subsection{NGC 2841}
\label{sectiongalacticthree}

The third galaxy in our analysis is NGC 2841, an unbarred spiral galaxy in the constellation of Ursa Major and approximately 46 million light-years away. This galaxy possesses a central spherical bulge, but the stellar disk component becomes prominent at small radial distances. The gas component is also present, but less prominent than the other two.
In the SPARC literature, this galaxy is mentioned and studied in Refs. \cite{2002ARA&A..40..263S,Famaey:2011kh,Li:2018tdo} and had also historically posed some challenges to MOND \cite{Begeman:1991iy,Gentile:2010xt}.

\begin{figure}\centering 
	\setlength\fboxrule{0in}\setlength\fboxsep{0.1in}\fcolorbox[HTML]{000000}{FFFFFF}{\includegraphics[ width=6.99in, height=8.728805970149253in,]{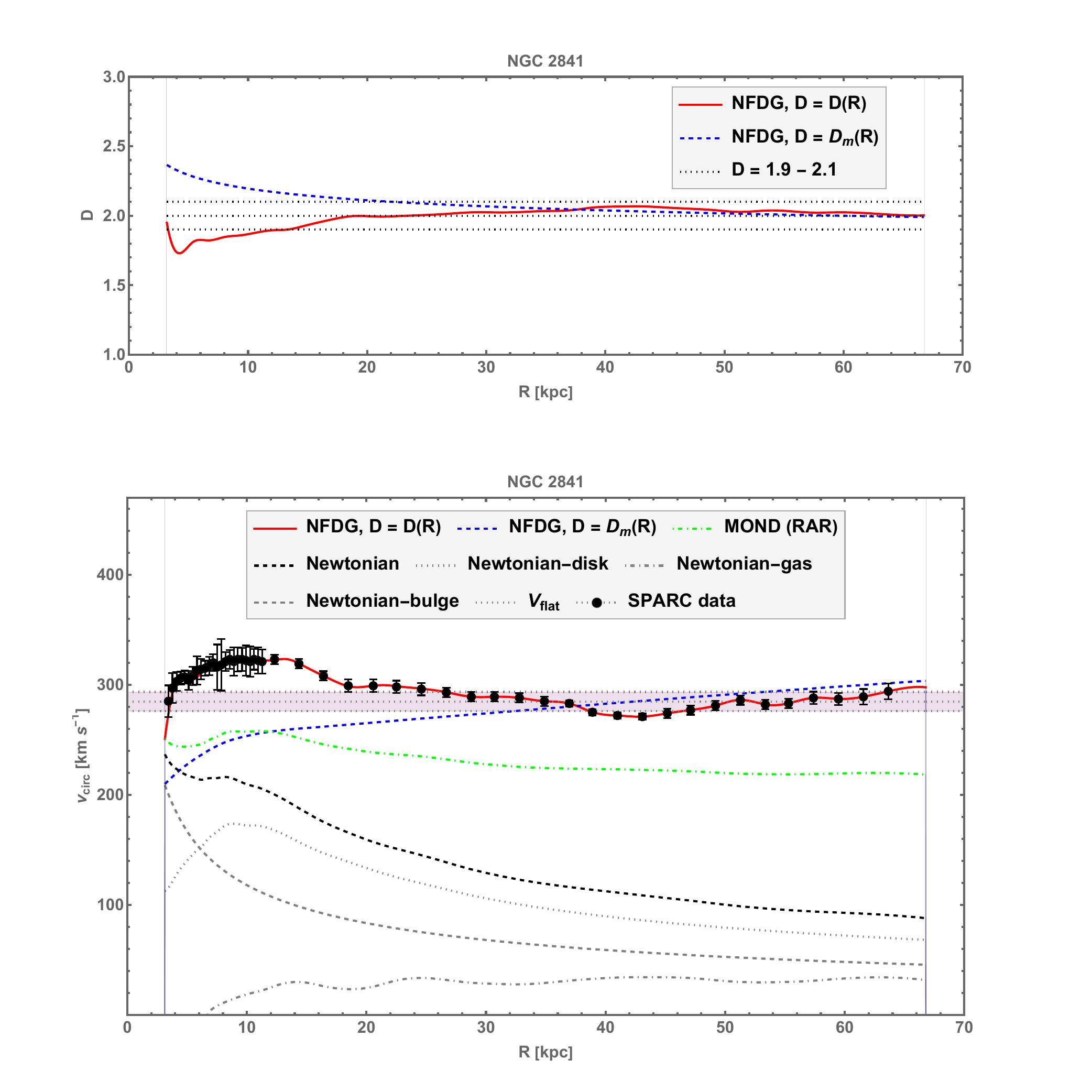}
	}
	\caption{NFDG results for NGC 2841.
		Top panel: NFDG variable dimension $D\left (R\right )$, based directly on SPARC data (red-solid curve), compared with NFDG mass-dimension $D_{m}\left(R\right)$ (blue-dashed curve), and fixed values $D =1.9-2.1$ (black-dotted lines). Bottom panel: NFDG rotation curves (circular velocity vs. radial distance) compared to the original SPARC data (black circles with error bars). The NFDG best fit for the variable dimension $D\left (R\right )$ is shown by the red-solid line, while the NFDG fit for the mass-dimension $D_{m}\left(R\right)$ is shown by the blue-dashed curve. Also shown: MOND prediction based on the general RAR (green, dot-dashed), Newtonian rotation curves (different components - gray lines, total - black dashed line), and asymptotic flat velocity band (horizontal gray band).}
	\label{figure:NGC2841}
\end{figure}

From SPARC data for this galaxy, we have the following: distance $D =\left (14.10 \pm 1.40\right)\ \mbox{Mpc}$, disk scale length $R_{d} =3.64\ \mbox{kpc} =1.12 \times 10^{20}\ \mbox{m}$, asymptotically flat rotation velocity $V_{f} =\left (284.8 \pm 8.6\right)\ \mbox{km/s}$. By integrating the SPARC mass distributions, we computed the following galactic masses: $M_{bulge} =9.72 \times 10^{39}\ \mbox{kg}$, $M_{disk} =1.54 \times 10^{41}\ \mbox{kg}$, \mbox{$M_{gas} =2.34 \times 10^{40}\ \mbox{kg}$,} and $M_{total} =1.87 \times 10^{41}\ \mbox{kg}$. The NFDG results for this galaxy are shown in Figure \ref{figure:NGC2841}, with the radial limits set at $R_{\min } =3.14 \ \mbox{kpc}
$ and $R_{\max } =66.8 \ \mbox{kpc}$ (vertical thin-gray lines in the figure).

As for the previous two galaxies, the variable dimension $D\left (R\right)$ is shown in the top panel of Figure \ref{figure:NGC2841} as a red-solid curve, with NFDG circular velocities computed using this $D\left (R\right)$ function producing again a perfect NFDG fit to the SPARC experimental data in the bottom panel of the same figure (red-solid curve). The mass-dimension function $D_{m}\left (R\right)$ can match the $D\left (R\right)$ only in the second half of the radial range, in the top panel of Figure \ref{figure:NGC2841}, so that the corresponding fit (blue-dashed curve) in the lower panel is not very effective in the lower range of radial distances. The MOND (RAR) fit without any parameter adjustment is also very ineffective, because this galaxy has notoriously very large values of the disk and bulge mass-to-light ratios \cite{Li:2018tdo}, which need to be used as free parameters to correct the fit based on Equation (\ref{eq4.3}). 

In general, the limited predictive capability of the mass-dimension Equation (\ref{eq3.4}), as seen in the case of NGC 2841 and of the other galaxies in this paper, is probably due to the approximations in the mass distributions functions  $\Sigma _{bulge}\left (R\right)$, $\Sigma _{disk}\left (R\right)$, $\Sigma _{gas}\left (R\right)$ for the three main components, which are obtained directly from the corresponding SPARC surface luminosity densities for each galaxy. In other words, these functions, obtained from the SPARC data, are usually quite ``smooth'' (especially for disk and gas components) and do not typically show the same features that are then found in the SPARC rotational curves at corresponding radial distances. Since the total surface mass distribution \linebreak $\Sigma_{tot} \left (R\right) =\Sigma _{bulge}\left (R\right) +\Sigma _{disk}\left (R\right)+\Sigma _{gas}\left (R\right)$ is the main input of Equation (\ref{eq3.4}), the resulting NFDG $D_{m}(R)$ fits in all our figures (blue-dashed curves) are also rather smooth and tend to match only the overall trend of the experimental data and not the finer details.

As already mentioned in Section \ref{sectiongalacticone}, the results obtained using the mass-dimension field Equation (\ref{eq3.4}) do not appear to be oversensitive to the choice of the free parameters $R_{0}$ and $M_{0}$ or to the values used in the initial condition in Equation (\ref{eq4.2}). This initial condition is necessary for the numerical solution of the mass-dimension differential equation. However, the value of $D\left(R_{data} \right)$, used in the same Equation (\ref{eq4.2}), could be left as an additional free parameter, thus avoiding setting the initial condition for $D_{m}\left (R\right)$ by using data obtained from the other dimension function $D\left(R \right)$.

Although our NFDG $D_{m}\left (R\right)$ fit is less effective for NGC 2841, both our methods suggest that this galaxy behaves as a $D \approx 2$ fractal structure for larger radial distances. Again, this might be due to the dominant stellar disk component, as in the previous case of NGC 3198.

\subsection{NGC 7814, NGC 6503, NGC 3741}
\label{sectiongalacticfour}

In this final subsection, we will revisit three rotationally-supported galaxies (NGC 7814, NGC 6503, NGC 3741), which were previously studied with NFDG methods in paper III \cite{Varieschi:2020hvp} and paper V \cite{Varieschi:2022mid}. In this way, we will be able to check that the changes and simplifications to our main NFDG calculations, outlined in Section \ref{sect:revised}, do not substantially affect the final results. We will also be able to apply our new analysis based on the field Equation (\ref{eq3.4}) to three additional cases. General descriptions of these three galaxies can be found in paper III \cite{Varieschi:2020hvp} and in Appendix B of paper V \cite{Varieschi:2022mid}. In this section, we will briefly compare our new results and figures with those in Appendix B of Ref. \cite{Varieschi:2022mid}.

\begin{figure}\centering 
	\setlength\fboxrule{0in}\setlength\fboxsep{0.1in}\fcolorbox[HTML]{000000}{FFFFFF}{\includegraphics[ width=6.99in, height=8.728805970149253in,]{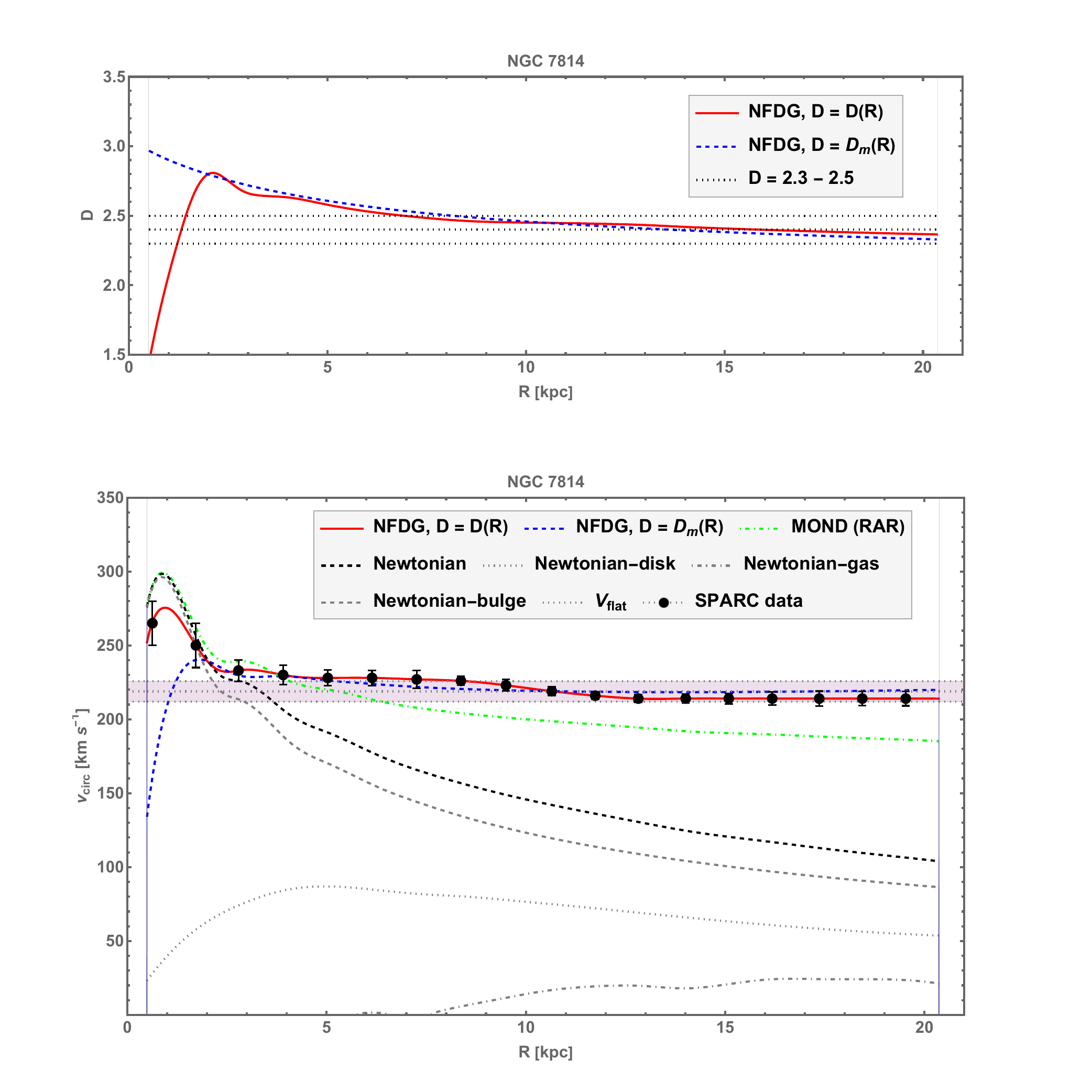}
	}
	\caption{NFDG results for NGC 7814.
		Top panel: NFDG variable dimension $D\left (R\right )$, based directly on SPARC data (red-solid curve), compared with NFDG mass-dimension $D_{m}\left(R\right)$ (blue-dashed curve), and fixed values $D =2.3-2.5$ (black-dotted lines). Bottom panel: NFDG rotation curves (circular velocity vs. radial distance) compared to the original SPARC data (black circles with error bars). The NFDG best fit for the variable dimension $D\left (R\right )$ is shown by the red-solid line, while the NFDG fit for the mass-dimension $D_{m}\left(R\right)$ is shown by the blue-dashed curve. Also shown: MOND prediction based on the general RAR (green, dot-dashed), Newtonian rotation curves (different components - gray lines, total - black dashed line), and asymptotic flat velocity band (horizontal gray band).}
	\label{figure:NGC7814}
\end{figure}

NGC 7814 is a spiral galaxy with a dominating central bulge and less conspicuous disk and gas components. Our new results are illustrated in Figure \ref{figure:NGC7814}, in the same way as the other figures in this paper. Comparing this new figure with Figure 5 in Appendix B of Ref. \cite{Varieschi:2022mid}, we can check that there are practically no differences between the old and new dimension functions $D\left (R\right)$ in the top panels and related fits to SPARC data in the bottom panels (red-solid curves in all figures). This shows that the simplified procedure detailed in Section \ref{sect:revised} does not change our main NFDG results. 

The new feature in Figure \ref{figure:NGC7814} is represented by the blue-dashed curves in both panels, describing the mass dimension $D_{m}\left (R\right)$ (top panel) and the related fit to SPARC data in the bottom panel. This new blue-dashed fit for NGC 7814 seems to work remarkably well and almost reproduces the red-solid fit at all radial distances. This confirms the validity of this alternative method, based on our fundamental field Equation (\ref{eq3.4}). NGC 7814 also presents similarities with the previously studied NGC 6946: they both have strong central bulges and prominent stellar disks at larger distances; so, they both end up having $D \approx 2.3-2.5$ in the outer radial range, where the circular velocities flatten out.

NGC 6503 is a field dwarf spiral galaxy with a dominating stellar disk and a less prominent gas component. Our new results are shown in Figure \ref{figure:NGC6503}. The comparison is with Figure 6 in Appendix B of Ref. \cite{Varieschi:2022mid}, and again, we find  no visible differences between the old and new dimension functions $D\left (R\right)$ in the top panels and the related fits to SPARC data in the bottom panels (red-solid curves in all figures), confirming the validity of our revised main computation. 

The new blue-dashed curves in Figure \ref{figure:NGC6503}, related to the mass dimension $D_{m}\left (R\right)$, are reasonably close to the previously described red-solid curves, although they cannot reproduce all the details of the experimental data. NGC 6503 is also similar to NGC 3198 studied in Section \ref{sectiongalactictwo}: they both are disk-dominated galaxies with $D \approx 2.0$ in the outer radial range, where the circular velocities flatten out.

\begin{figure}\centering 
	\setlength\fboxrule{0in}\setlength\fboxsep{0.1in}\fcolorbox[HTML]{000000}{FFFFFF}{\includegraphics[ width=6.99in, height=8.728805970149253in,]{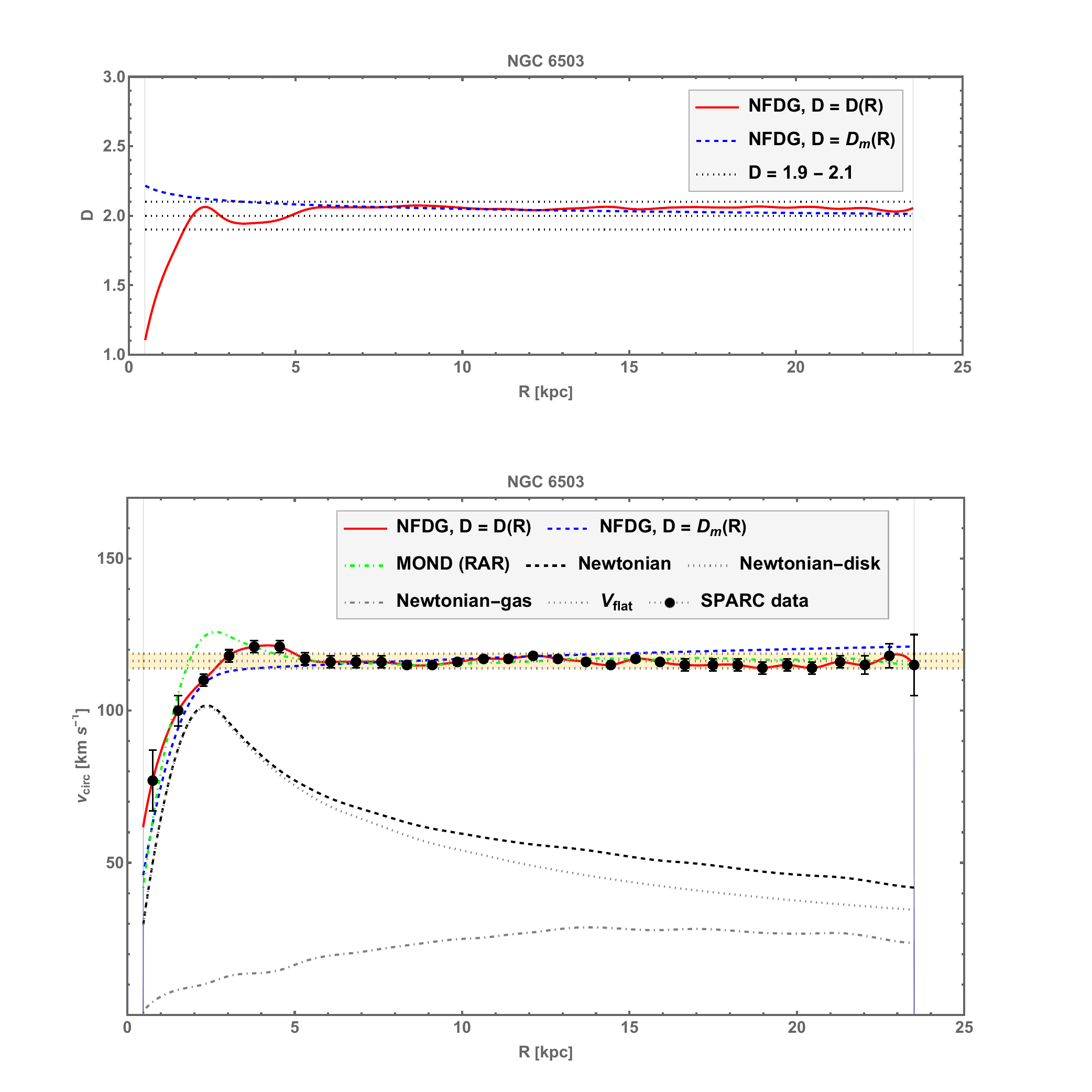}
	}
	\caption{NFDG results for NGC 6503.
		Top panel: NFDG variable dimension $D\left (R\right )$, based directly on SPARC data (red-solid curve), compared with NFDG mass-dimension $D_{m}\left(R\right)$ (blue-dashed curve), and fixed values $D =1.9-2.1$ (black-dotted lines). Bottom panel: NFDG rotation curves (circular velocity vs. radial distance) compared to the original SPARC data (black circles with error bars). The NFDG best fit for the variable dimension $D\left (R\right )$ is shown by the red-solid line, while the NFDG fit for the mass-dimension $D_{m}\left(R\right)$ is shown by the blue-dashed curve. Also shown: MOND prediction based on the general RAR (green, dot-dashed), Newtonian rotation curves (different components - gray lines, total - black dashed line), and asymptotic flat velocity band (horizontal gray band).}
	\label{figure:NGC6503}
\end{figure}

Finally, NGC 3741 is an irregular galaxy with a dominating gas component and less prominent stellar disk. Our new results are shown in Figure \ref{figure:NGC3741} and are compared with Figure 7 in Appendix B in Ref. \cite{Varieschi:2022mid}. Again, almost no visible differences can be seen between the respective red-solid curves of the new and old figures, while the new blue-dashed curves in Figure \ref{figure:NGC3741} are reasonably effective in fitting the SPARC data, although they cannot reproduce perfectly all the finer details. NGC 3741 is not similar to any of the other galaxies studied in this paper, since it has a dominating gas component, which might be responsible for the small dimension value ($D \approx 1.4-1.5$) in the outer radial range.

\begin{figure}\centering 
	\setlength\fboxrule{0in}\setlength\fboxsep{0.1in}\fcolorbox[HTML]{000000}{FFFFFF}{\includegraphics[ width=6.99in, height=8.728805970149253in,]{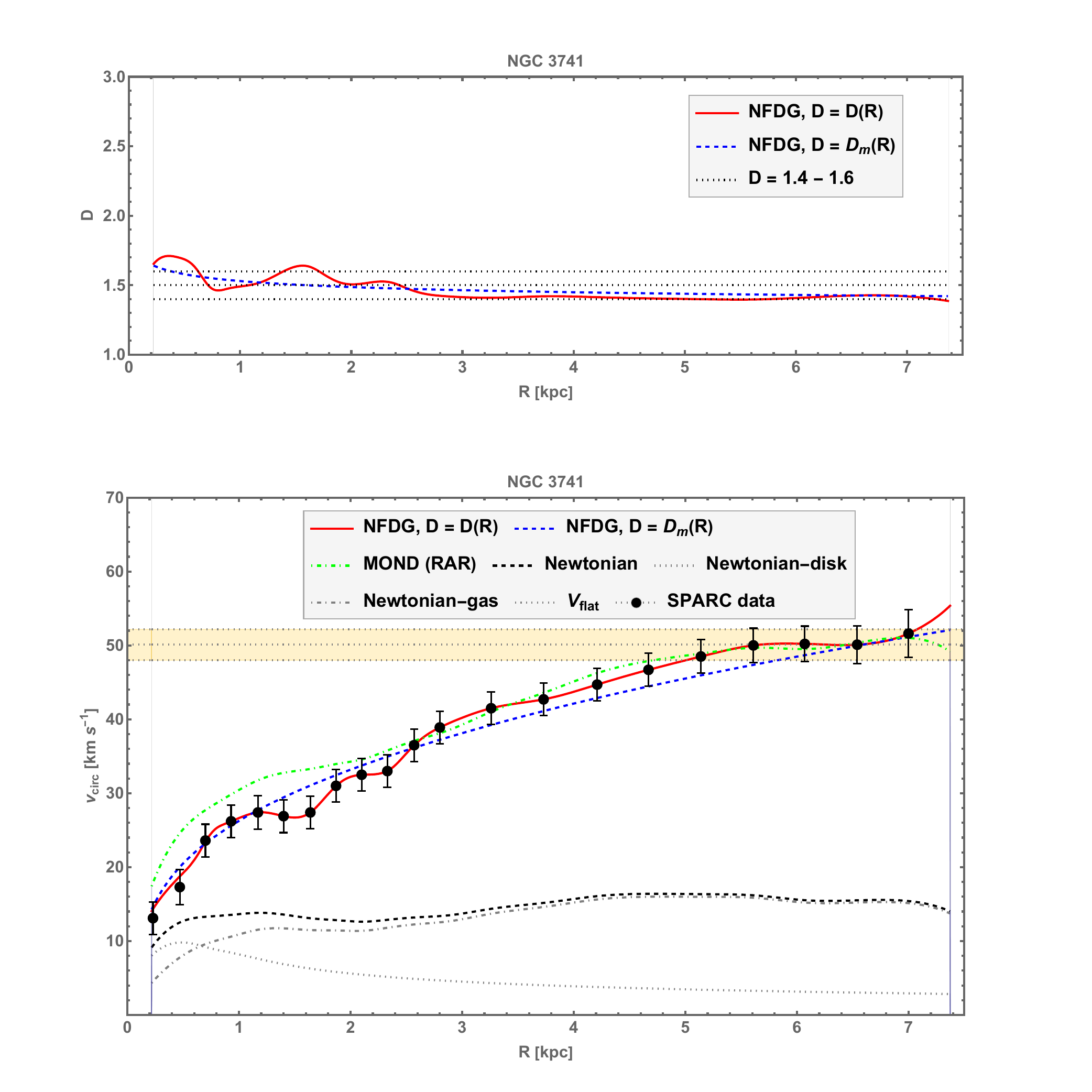}
	}
	\caption{NFDG results for NGC 3741.
		Top panel: NFDG variable dimension $D\left (R\right )$, based directly on SPARC data (red-solid curve), compared with NFDG mass-dimension $D_{m}\left(R\right)$ (blue-dashed curve), and fixed values $D =1.4-1.6$ (black-dotted lines). Bottom panel: NFDG rotation curves (circular velocity vs. radial distance) compared to the original SPARC data (black circles with error bars). The NFDG best fit for the variable dimension $D\left (R\right )$ is shown by the red-solid line, while the NFDG fit for the mass-dimension $D_{m}\left(R\right)$ is shown by the blue-dashed curve. Also shown: MOND prediction based on the general RAR (green, dot-dashed), Newtonian rotation curves (different components - gray lines, total - black dashed line), and asymptotic flat velocity band (horizontal gray band).}
	\label{figure:NGC3741}
\end{figure}

\section{\label{sect:conclusion}Conclusions}
In this work, we improved and expanded our NFDG model and related computations of galactic dynamics, and we applied them to three new galaxies from the SPARC catalog. Our main NFDG calculations, based on the variable dimension function $D\left (R\right)$, were streamlined in order to reduce the computing times for each galaxy being analyzed. These improvements and simplifications were proven to be effective. We are also convinced that this latest version of the NFDG computation is mathematically correct and that our final results are noteworthy.

In addition, we introduced a mass-dimension field equation in order to determine an alternative dimension function $D_{m}\left (R\right)$, obtained from this fundamental equation. The new mass-dimension function should be equivalent to the previous one, i.e., $D_{m}\left (R\right) \approx D\left (R\right)$, but differences might be present due to the separate computations leading to these functions. We applied both methods to three new galaxies selected from the SPARC catalog (NGC 6946, NGC 3198, NGC 2841), as well as three previously studied galaxies from the same catalog (NGC 7814, NGC 6503, NGC 3741).

For all these galaxies, our standard NFDG analysis remains fully effective in describing the rotational curves without any DM contribution, while our new method based on the mass-dimension field equation does not achieve the same level of accuracy in matching the experimental data; nevertheless, it confirms the validity of our approach. 

All these methods will be improved in future work on the subject, and more galaxies will be added to the NFDG catalog. The fractional-dimension gravity approach in astrophysics and cosmology will continue to be relevant primarily as a theoretical framework used to address the dark matter problem, as well as other issues in the standard $\Lambda CDM$ cosmological model, including the accelerated expansion of the universe and the dark energy problem.

\begin{acknowledgments}This study was supported by the Seaver College of Science and Engineering, Loyola Marymount University---Los Angeles. The author also wishes to acknowledge Dr. Federico Lelli for sharing the latest SPARC data and other information, Mr. Isaiah Tyler---LMU student---for helping with the computations of NGC 3198, and the anonymous reviewers for useful comments and~suggestions.
\end{acknowledgments}

\section*{Data Availability}

The data underlying this article will be shared on reasonable request to the corresponding author.

\end{document}